\documentclass[journal]{IEEEtran}
\usepackage{blindtext,booktabs}
\usepackage{caption}
\usepackage{xcolor}
\usepackage[T1]{fontenc}

\usepackage{cite}

\ifCLASSINFOpdf
\usepackage[pdftex]{graphicx}
\else
\usepackage[dvips]{graphicx}
\fi
\usepackage[cmex10]{amsmath}
\usepackage{algorithmic}
\usepackage[ruled,lined,boxed,commentsnumbered]{algorithm2e}

\usepackage{array}

\usepackage{mdwmath}
\usepackage{mdwtab}

\usepackage{eqparbox}

\usepackage[caption=false,font=footnotesize]{subfig}
\begin{document}
%
\title{Dynamic Placement of VNF Chains for Proactive Caching in Mobile Edge Networks}
%
%
%

\author{\IEEEauthorblockN{Gao~Zheng,\thanks{Partially funded by the EC H2020-ICT-2014-2 project 5G NORMA (\texttt{www.5gnorma.5g-ppp.eu})}~Anthony~Tsiopoulos,~Vasilis~Friderikos}\\        
\IEEEauthorblockA{Centre for Telecommunications Research, King's College London, London
WC2R 2LS, England
\\ E-mail: \{gao.zheng, anthony.tsiopoulos, vasilis.friderikos\}@kcl.ac.uk}
}

\maketitle

\begin{abstract}
Notwithstanding the significant research effort Network Function Virtualization (NFV) architectures received over the last few years little attention has been placed on optimizing proactive caching when considering it as a service chain. Since caching of popular content is envisioned to be one of the key technologies in emerging 5G networks to increase network efficiency and overall end user perceived quality of service we explicitly consider in this paper the interplay and subsequent optimization of caching based VNF service chains. To this end, we detail a novel mathematical programming framework tailored to VNF caching chains and detail also a scale-free heuristic to provide competitive solutions for large network instances since the problem itself can be seen as a variant of the classical $NP$-hard Uncapacitated Facility Location (UFL) problem. A wide set of numerical investigations are presented for characterizing  the  attainable  system  performance of the proposed schemes. 

\end{abstract}


%
\IEEEpeerreviewmaketitle

\section{Introduction}
\begin{figure}
\includegraphics[width=0.9\columnwidth]{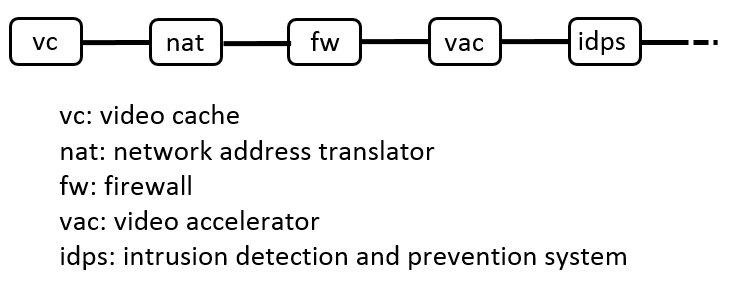}
\caption{Caching as a VNF chain.}
\label{fig:example_chain}
\end{figure}

\IEEEPARstart{I}{t} is well accepted that current mobile network architectures suffer from insufficient scalability and flexibility to quickly accommodate new services and ability to embrace vertical industries \cite{VNF state of art}. To address these challenges, applying software define networking (SDN)\cite{SDN} principles in emerging architectures towards 5G networks is gaining significant momentum recently\cite{Scalable and Flexible cellular core network Architecture}. This goes hand-in-hand With the heavily studied now network function virtualization (NFV) \cite{NFV} architectures, that together with SDN, can be considered as the two enablers towards flexible wireless networks, where full virtualization and efficient network slicing according to the needs of different tenants can be implemented. An SDN/NFV-enabled network is in essence able to decouple network functions (NFs) from the underlying physical devices, thereby, NFs can be virtualized, creating the so-called virtual network functions (VNFs). The benefit is that VNFs can be flexibly controlled/assigned/moved within the network using Virtual machines or (docker) containers. In NFV framework, an end-to-end network service (e.g., rich voice/data) is described by an VNF forwarding graph, where a number of VNFs (possibly distributed in various physical nodes in the network) need to be visited in certain predefined order\cite{VNF Chaining basic}. To be more precise, the sequenced VNFs of a service request form a service chaining as the service flow passes through an ingress or egress point in a virtual network device. An illustrative example of such service chain is shown in figure \ref{fig:example_chain}, where caching is considered as one of the  VNFs\footnote{The terms VNF and NF are used interchangeably in the rest of the paper, except where differentiation is required.} that constitute the overall service chain; these VNFs might be located in different nodes in the network. Our aim is to consider caching and the other possible VNFs that might be required for the service in an integrated manner in order to increase network efficiency. 

Undoubtedly, among different VNFs, it is expected that caching would emerge as one of the potential key network elements to be supported in emerging and future wireless/mobile networks.  Viral and popular video streams dominate aggregate mobile Internet traffic\footnote{Mobile video traffic accounts for 60 percent of total mobile data traffic according to  the CISCO Global Mobile Data Traffic Forecast Update that has been released in February 2017.} and it is an application well suited to various different caching strategies. In that respect, caching of popular content deserves paying a special attention in terms of VNF hosting location and chaining. This is because in the most general case, a cached content must be visited before other VNFs can be applied and this service flow might originate from different possible network locations depending on the caching strategy. Hence, the service does need to reach a gateway node but can originate at a node that host the required cached content (which can be topologically close to the end user). Therefore, the location of caches in a VNF service chain, greatly affects the overall VNF chain orchestration as well as the aggregate traffic dynamics in the network, since links of higher aggregation (deeper in the network) can reduce their utilization levels. However, efficient caching in mobile networks can be deemed as a highly challenging task since the optimality of the cache locations are dependent on the movement/mobility patterns of the users. Notably, to significantly reduce access delays to highly popular content caching content close to the end user without considering the effect of mobility might lead to degradation of performance. In this case, caching popular content closer to the end user might inevitably require more frequently changes of the cache location to keep providing optimal performance. As a result, the caching location and the associated VNF chaining need to be jointly considered to avoid sub-optimal cases, especially under congestion episodes where performance can be significantly affected. To summarize, the focus and motivation of the paper is on enhancing proactive caching  policies by taking into account the whole VNF chain.
\section{Motivation and Illustrative Examples}
\begin{figure}
\includegraphics[width=0.9\columnwidth,trim=0cm 9cm 15cm 0.0cm]{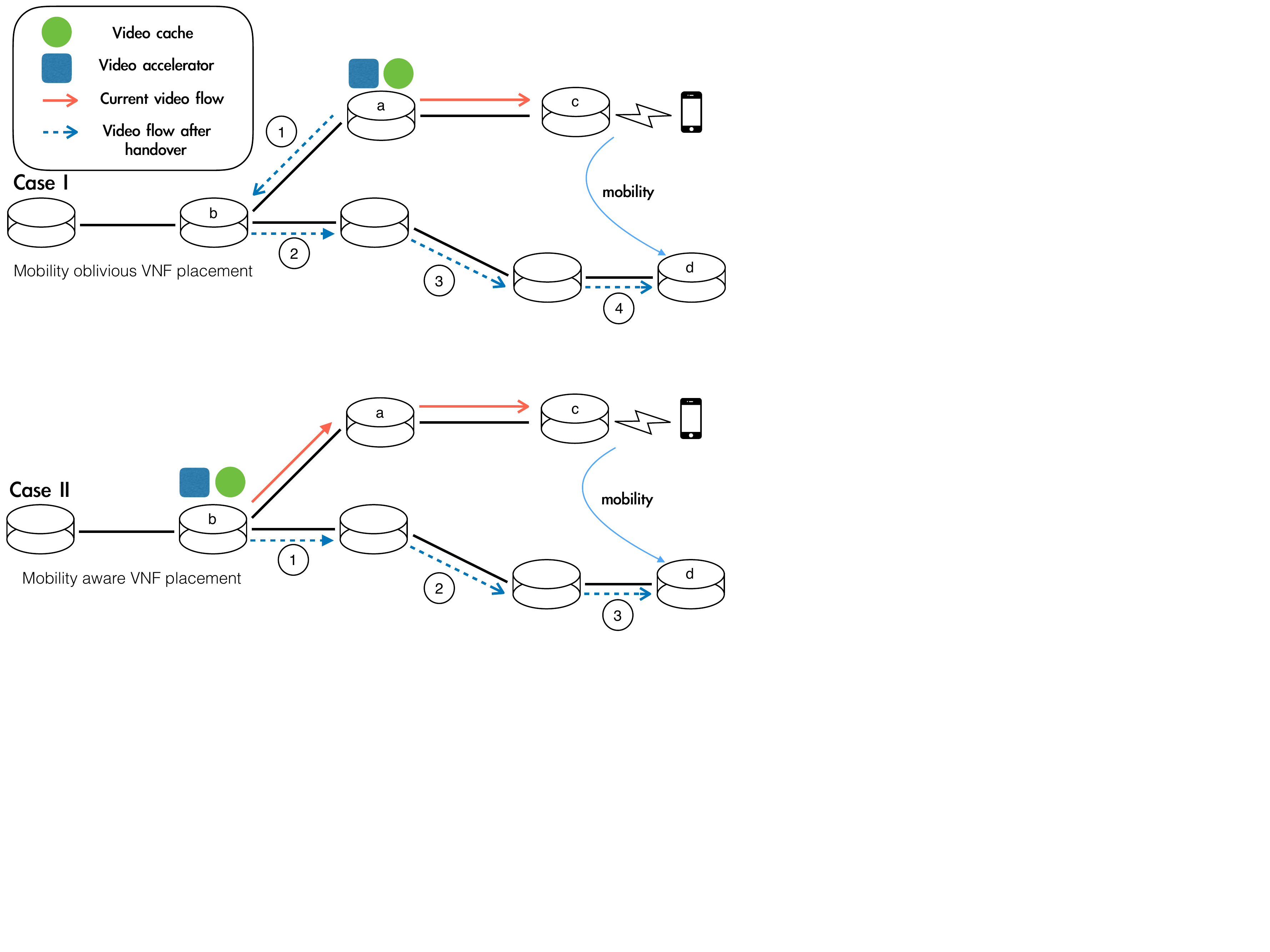}
\caption{Effect of mobility on the joint caching VNF chaining problem.}
\label{fig:mobility_chain}
\end{figure}

\begin{figure}
\includegraphics[width=0.9\columnwidth,trim=0cm 17cm 15cm 0.0cm]{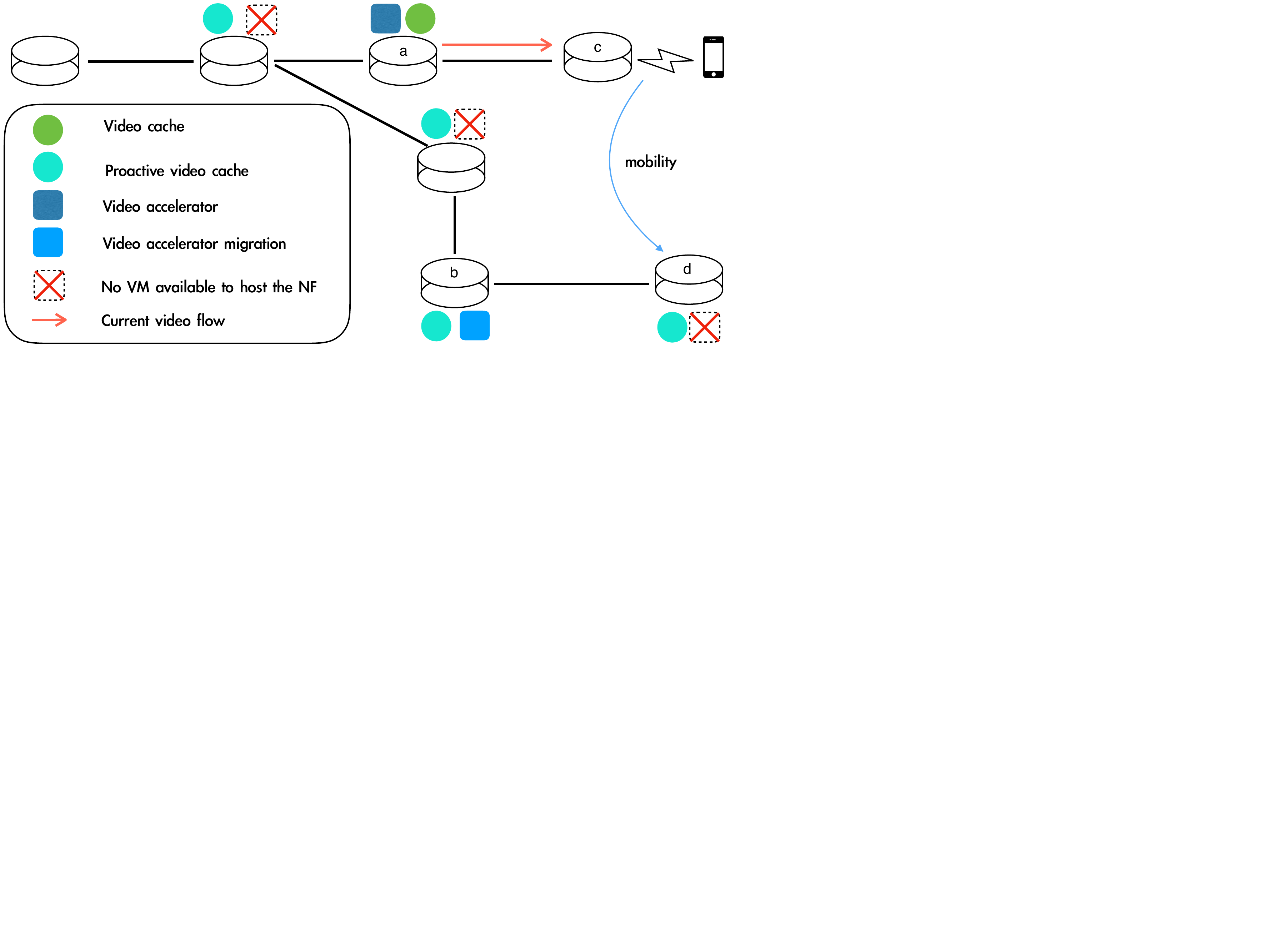}
\caption{Limited availability of resources (in terms of Virtual Machines for example) in the candidate pro-active caching locations to host the required VNFs for the service.}
\label{fig:noVM} 
\end{figure}

\begin{figure}
\includegraphics[width=0.9\columnwidth,trim=0cm 17cm 15cm 0.0cm]{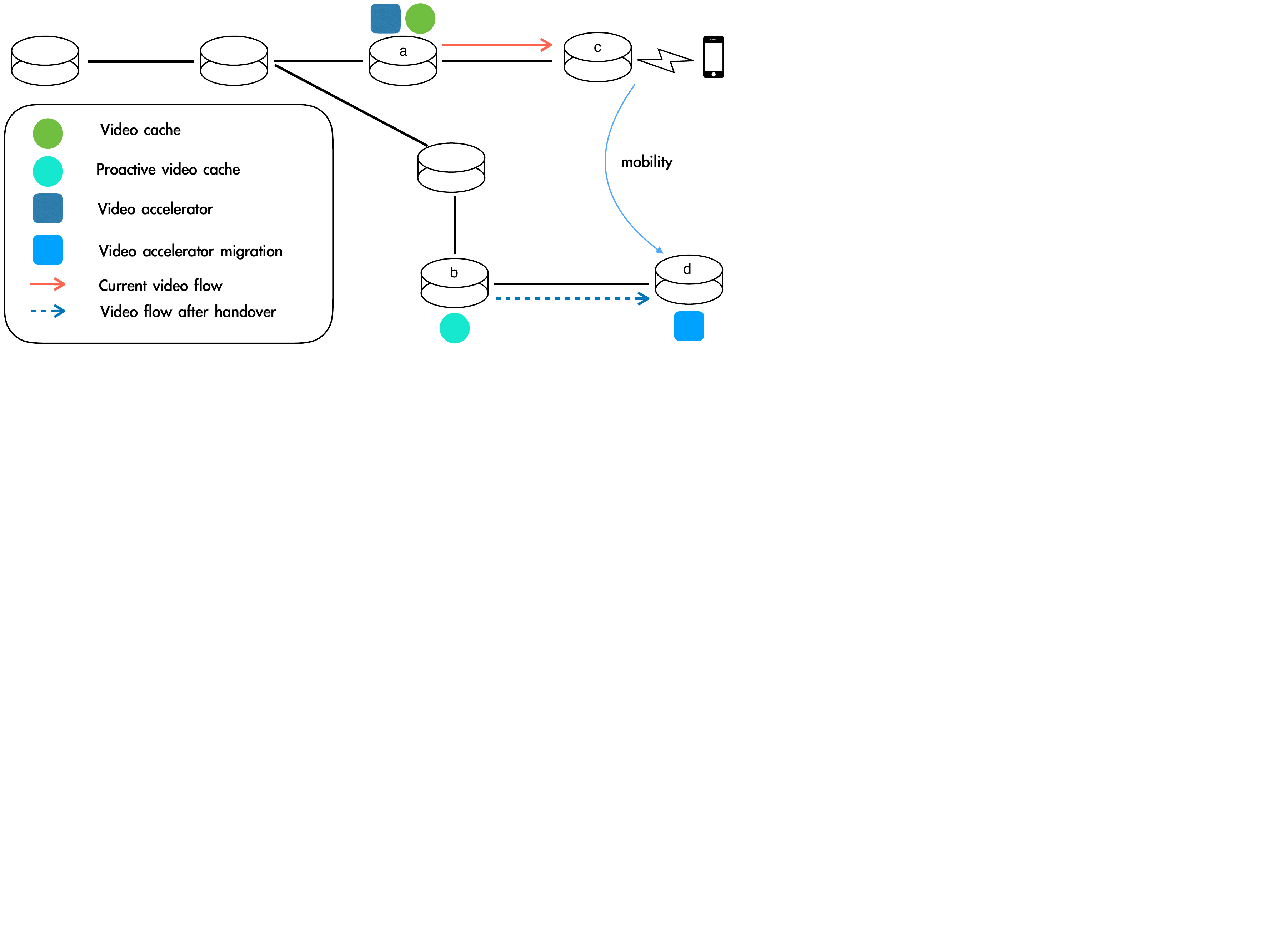}
\caption{An optimal VNF chain NF are located in different nodes in the network.}
\label{fig:opt_cache_chain}
\end{figure}

In this paper, we propose a Proactive caching-chaining (PCC) scheme to enhance the mobility support of SDN-enabled/NFV service chaining in mobile networks. To motivate the research we discuss illustrative examples of the cross issues between caching and VNF chaining that aim to shed further light on some of the key challenges. To start with, Figure \ref{fig:mobility_chain} shows the case of a service with two NF where the first one is caching and the other one is assumed to be a video acceleration network function. As can be seen from the figure, Case I entails a  sub-optimal allocation when mobility is also taken into account. Case II shows a more suitable NF location where after the mobility event the cache and chain location is topologically closer to the end user; in Case II the NFs are located 3 hops away from the end user after the mobility event whereas in Case I, which a mobility oblivious allocation the NFs are located 4 hops away. Figure \ref{fig:noVM} shows the case where VNF chaining and pro-active caching take place independently. The figure shows potential pro-active caching locations but not in all of those pre-selected locations from the caching algorithm it is possible to host the other NFs due to numerous reasons such
as for example reservation policies, placement based on affinity and/or anti-affinity rules and overall resource usage of the virtual machines \cite{affinity_NF}; for example only in one of those locations the two NF can be co-located (node b). Furthermore, as shown in figure \ref{fig:opt_cache_chain} the optimal location of caching and the other NF in the service chain might be different; in the figure shown the optimal location of caching is in node (b) whereas the NF for video acceleration is located at node (d). It is therefore important firstly to consider the issue of caching and service chaining in a holistic integrated manner and secondly to optimize the location and chaining of the different NF in order to increase overall network performance. 

Based on the above discussion, the proposed scheme proactively performs caching and VNF chaining so that overall network performance is optimized whilst end user receive their requests seamlessly. Notably, we take VNF chaining allocation and proactive caching as a joint problem and formulate it as a Integer Linear Programming (ILP) problem that minimize the combine cost of VNF placement, chaining and routing. We also investigate the performance obtained of a proposed scale-free heuristic algorithm since the problem resembles the $NP$-hard UFL optimization problem.

In summary, we hereafter make the following key contributions,
 We firstly, propose a novel VNF chaining placement scheme, namely, proactive caching-chaining (PCC) that improves the mobility support for the up coming SDN-enabled/NFV network framework.
 Furthermore, we model and formulate the VNF chaining problem for proactive caching to obtain optimal routing and placement cost and based on that we devise a scalable heuristic approach and evaluate the performance of the system.

\section{Previous Research Work}
The overall logical architecture of the so-called VNF Management and Orchestration (MANO) architecture has been mainly an industry-lead initiative and has been defined within ETSI \cite{NFV}. An example of a VNF orchestrator, which is called \emph{Stratos} is presented in\cite{Stratos} and is built on top of a \emph{Floodlight}\footnote{\texttt{www.projectfloodlight.org}} controller. The work in \cite{Split/Merge} can be considered as another effort to provide orchestration between virtualized NFs especially with  emphasis on issues such as Virtual Machine (VM) migration and split/merging of service flows. An overview of the challenges emerging in virtual network function scheduling is presented in \cite{Riera}; in this paper the authors explain the application of SDN and NFV technologies with emphasis on backbone networks. 
In terms of caching there has been recently a significant amount of work. A caching scheme suitable for mobile networks that takes into account user mobility has been proposed in  \cite{op_caching} where the idea is to  predict  the mobility pattern of users and opportunistically cache content along the predicted path of users. A scheme that pro-actively cache content using transportation and focusing on video content has been presented in \cite{caching_transportation}. The idea is to utilize the almost deterministic mobility of users in transportation systems such as trains to proactively cache popular content that the users might request upon their arrival. The ideas on proactive caching in this paper
 resemble more closely the work in \cite{PCWR, Efficient proactive caching for support seamless mobility} which propose a set of mobility-aware caching schemes. 
 
 However, none of previous research works make caching decisions on a view of the whole service chain. To the best of our knowledge this is the first work to consider in an explicit and integrated manner proactive caching as part of a VNF chain. In most practical cases, this simple cache moving could lead to inefficient routing of a mobile user to receive a service. Fig 1 gives an example of the inefficient routing problem where firewall as a NF must also be visited and only cache is moved \footnote{NF movement in this paper refers to any approach that occurs the change of the function's location. (e.g., proactive caching)}. It is apparent that, in order to improve the mobility support of SDN-enabled networking, other NFs on a same VNF service chain must also be moved, with the decision of caching. A close related work can be found  in \cite{specifying and placing chains of virtual network functions} which aims to assign VNFs into given SDN-enabled networks. However, it does not take routing and location of VNFs into consideration.

\section{Network Modeling and Proactive chaining with caching}

A mobile network is modeled as an undirected  graph $G=(\mathbf{N},\mathbf{E})$, where $\mathbf{N}$ denotes the set of nodes in the network and $E$ denotes the set of links in the network. By $\mathbf{F}$, we denote the set of NFs and $f_{i}$ represents the specific NF$_{i}$. Each $f_{i}$, if activated, consumes/requires some physical resources (i.e., CPU cycles, DRAM memory). We uniformly describe these resource requirements as a single column matrix $u_{i}$, meanwhile, the amount of available resources of node $k$, which is able to host VNFs, is denoted using the single column matrix $U_{k}$.

The term "chain" in the so-called service chaining represents the different middleboxes that the service should traverse, with a specific order, across the network using software provisioning. This is the case under the proposed NFV architecture, where new services and/or network slices can be instantiated as software-only, running on commodity hardware on top of virtual machines or containers. To provide a service request $r\in \mathbf{R}$ (with $\mathbf{R}$ we denote the set of requests\footnote{VNF chain deployment in this paper is for per user request. Nevertheless, coarser grain can be applied without change to our model by referring the request as a set of traffic. This is based on the granularity of VNF deployment and usage.}) for a mobile user and/or tenant, a network function forwarding graph (VNF-FG)\cite{VNF-FG} needs to access a set of corresponding NFs that are visited in a pre-defined order (which the VNF orchestrator should preserve). In this paper, we consider the form of service request $r$ as the set $r=\{f_{1}, f_{2}, \cdots, f_{i}\}$ where the sequence express the visiting order of the different network functions. The proposed optimization scheme provides a batch processing based service that the requests are handled in batches, such that the number of requests processed in each batch is $|\mathbf{R}|$. For modeling simplification reasons, the corresponding relationship of a NF and its order in a request can be represented by a binary matrix $V_{ril}$ as follows,

\begin{equation}
V_{ril} = \left \{
\begin{array}{rl}
1 & \text{if the $l^{th}$ NF of request $r$ is NF$_{i}$}. \\
0 & \text{otherwise}.
\end{array} \right.
\end{equation}

Hereafter, we consider the scenario where a mobile user and/or tenant connected to node $o$ and requesting $\mathbf{R}$ services. As presented in Figure 2, caching as a NF, is the head of a service request chain and it is denoted as $f_{0}$. We define a candidate node set $\mathbf{K}\subseteq \mathbf{N}$ that consist of the potential candidate nodes of hosting NFs. By $\mathbf{D}$, we define a set of potential destinations that mobile users might move due to their inherent mobility. Using historical data available to mobile network providers it is feasible to  estimate such probabilities of end users moving from their current  location to an adjacent candidate destination node $d$. We denote this probability of changing their serving access router with $\rho_{d}$. As eluded, we assume that $\rho_{d}$ is predefined by using available historical data from operators so this assumption can be deemed as realistic due to vast available data which can provide accurate  characterization of user mobility patterns. With known candidate cache locations, which can be done using for example a proactive caching technique such as PCWR\cite{PCWR}), PCC aims to proactively place network functions $f_{i}\in \mathbf{F}$ into the set of nodes $\mathbf{K}$. To be more precise, we define by $\mathbf{S_{r}}$ to be the set of initiating nodes (i.e., proactive caching locations) of a service chain $r$, with  $\mathbf{H}$ denoting the set of $\mathbf{S_{r}}$. Given $\mathbf{H}$ and $\mathbf{D}$, the proposed scheme returns the optimal proactive allocation of the NFs that minimizes the joint cost of routing, location and chaining.

\subsection{Proactive chaining-caching problem}
Based on the previously described network settings we define the following binary decision variables,

\begin{equation}
x_{ri} ^{k} = \left \{
\begin{array}{rl}
1 & \text{if NF$_{i}$ is placed at $k$ for request $r$}. \\
0 & \text{otherwise}.
\end{array} \right.
\end{equation}

\begin{equation}
y_{ri} ^{ksd} = \left \{
\begin{array}{rl}
1 & \text{if NF$_{i}$ of request $r$ with head $s$ and} \\
&\text{destination $d$ is visited from $k$}. \\
0 & \text{otherwise}.
\end{array} \right.
\end{equation}

To linearize the optimization problem, we introduce an auxiliary variable $z_{rij}^{kmsd}$ which holds the value of product $y_{ri}^{ksd}y_{rj}^{msd}$ defined as follows,

\begin{equation}
z_{rij} ^{kmsd} = \left \{
\begin{array}{rl}
1 & \text{if request $r$ with head $s$ and destination $d$} \\
&\text{visits NF$_i$ at node $k$ and NF$_j$ at node $m$}. \\
0 & \text{otherwise}.
\end{array} \right.
\end{equation}

The optimal VNF location and chaining for the proactive caching problem is defined as the following non-linear integer optimization problem where the first term of the objective function is the placement cost of hosting VNFs at a node and the rest terms reflect the accumulative routing cost of each hop on the VNF-FG of a requested chain:

\begin{equation}
\begin{split}
\underset{x_{ri}^{k},y_{ri}^{ksd}}{\min}\!\! \: \sum_{r \in \mathbf{R}}\!\sum_{k \in \mathbf{K}}\!\sum_{i \in \mathbf{F}} C_{i}^{k} x_{ri}^{k}\!\!+\!\!\sum_{r \in \mathbf{R}}\sum_{s \in \mathbf{S_{r}}}\!\sum_{d \in \mathbf{D}} \sum_{k \in \mathbf{K}}\sum_{i \in \mathbf{F}} \rho_{d} P_{sk} V_{ri1} y_{ri}^{ksd}\\+ \sum_{r \in \mathbf{R}}\sum_{s \in \mathbf{S_{r}}}\sum_{d \in \mathbf{D}}\!\sum_{k,m \in \mathbf{K}}\sum_{i,j \in \mathbf{F}} \sum_{l=1}^{L-1} \rho_{d} P_{km}V_{ril} V_{rj(l+1)} z_{rij}^{kmsd}\\ + \sum_{r \in \mathbf{R}}\sum_{s \in \mathbf{S_{r}}} \sum_{d \in \mathbf{D}}\sum_{k \in \mathbf{K}}\sum_{i \in \mathbf{F}} \rho_{d} P_{kd} V_{riL} y_{ri}^{ksd}
\end{split}
\end{equation}
\label{eq:con1}
\begin{IEEEeqnarray}{cl}
\text{S.t.} \: \: 
  \sum_{r \in \mathbf{R}} \sum_{i \in \mathbf{F}}  u_{i} x_{ri}^{k} \leq U_{k} , \forall k \in \mathbf{K} \IEEEyessubnumber \\
\sum_{r \in \mathbf{R}}\sum_{d \in \mathbf{D}} \sum_{i \in \mathbf{F}}  \lambda_{r}V_{ri1} y_{ri}^{ksd} \leq \Lambda_{sk} , \forall r\! \in \!\mathbf{R}, k\!\! \in\! \mathbf{K}, s\! \in\! \mathbf{S_{r}} \IEEEyessubnumber \\
\sum_{r \in \mathbf{R}}\sum_{s \in \mathbf{S_{r}}}\sum_{d \in \mathbf{D}} \sum_{i,j \in \mathbf{F}} \sum_{l=1}^{L-1} \lambda_{r}V_{ril}V_{rj(l+1)} y_{ri}^{ksd}y_{rj}^{msd} \leq \Lambda_{km} \nonumber,\\ \forall k,m \in \mathbf{K} \IEEEyessubnumber \\
\sum_{r \in \mathbf{R}}\sum_{s \in \mathbf{S_{r}}} \sum_{i \in \mathbf{F}}  \lambda_{r}V_{riL} y_{ri}^{ksd} \leq \Lambda_{kd} , \forall k \in \mathbf{K}, d \in \mathbf{D} \IEEEyessubnumber \\
\label{eq:con2}
  \sum_{k \in \mathbf{K}}\sum_{i \in \mathbf{F}}V_{ril}y_{ri}^{ksd} \geq 1, \: \forall r \in \mathbf{R}, s \in \mathbf{S_{r}}, d \in \mathbf{D}, \nonumber \\l=1, \ldots L  \IEEEyessubnumber \\
\label{eq:con3}
 y_{ri}^{ksd}-x_{ri}^{k}\leq 0, \forall r\in \mathbf{R}, i\in \mathbf{F}, k \in \mathbf{K}, s \in \mathbf{S_{r}}, d \in \mathbf{D} \IEEEyessubnumber \\
\label{eq:con4}
z_{rij}^{kmsd} \leq y_{ri}^{ksd}, \forall r\in \mathbf{R}, i\in \mathbf{F}, k \in \mathbf{K}, s \in \mathbf{S_{r}}, d \in \mathbf{D} \IEEEyessubnumber \\
 \label{eq:con5}
z_{rij}^{kmsd} \leq y_{rj}^{msd}, \forall r\in \mathbf{R}, j\in \mathbf{F}, m \in \mathbf{K}, s \in \mathbf{S_{r}}, d \in \mathbf{D} \IEEEyessubnumber \\
 \label{eq:con6}
z_{rij}^{kmsd} \geq y_{ri}^{ksd}+y_{rj}^{msd}-1, \forall r\in \mathbf{R}, i,j\in \mathbf{F}, k,m \in \mathbf{K}, \nonumber \\ s \in \mathbf{S_{r}}, d \in \mathbf{D} \IEEEyessubnumber \\
 \label{eq:con7}
  x_{ri}^{k} \in \{0,1\}, \:\:\: \forall i \in \mathbf{F}, k \in \mathbf{K}\IEEEyessubnumber \\
\label{eq:con8}
  y_{ri}^{ksd} \in \{0,1\}, \:\:\: \forall r \in \mathbf{R}, i \in \mathbf{F}, k \in \mathbf{K}, s \in \mathbf{S_{r}}, d \in \mathbf{D} \IEEEyessubnumber \\
\label{eq:con9}
  z_{rij}^{kmsd} \in \{0,1\}, \:\:\: \forall r \in \mathbf{R}, i,j \in \mathbf{F}, k,m \in \mathbf{K}, \nonumber\\s \in \mathbf{S_{r}}, d \in \mathbf{D} \IEEEyessubnumber
\end{IEEEeqnarray}
where $C_{i}^{k}$ is the cost of placing NF$_{i}$ at $k$. While $P_{sk}$, $P_{km}$ and $P_{kd}$ are the shortest path routing costs between the candidate nodes. Accordingly, $\Lambda_{sk}$, $\Lambda_{km}$ and $\Lambda_{kd}$ are the remaining link capacities of a path (i.e., the bottleneck link capacity) and notice that $\Lambda_{km}$ can be seen as flow rate tolerant of a node when $k=m$. $\lambda_{r}$ denotes the flow rate requirement of request $r$. Constraint (5a) 
is the VNF processing capacity constraint which takes into account the CPU cycles associated with the Virtual Machine(VM) allocated to a VNF and the memory capacity for a specific VNF. (5b)-(5d) are the QoS constraints related to the service chain such that the requests can be properly assigned based on the flow rate requirements and the link capacity. (5e) enforces that each NF in a requested chain must be visited at least once. (5f) is a binding constraint that insures the availability of a NF at a node is valid only when the NF is hosted at the node. (5g)-(5i) are binding constraints that insure $z_{rij}^{kmsd}$ taking the same value as product $y_{ri}^{ksd}y_{rj}^{msd}$.

\section{A Scale Free Heuristic Approach}
The PCC problem falls within the family of $\mathcal{NP}$-hard problems since it resembles the UFL problem and as a result heuristics becomes the only viable option of finding competitive feasible solutions for real time operation. Therefore, a heuristics algorithm named, Probability-prior proactive caching-chaining (PPCC) is proposed and is detailed in the pseudocode Algorithm \uppercase\expandafter{\romannumeral1} below. The main philosophy of the proposed PPCC heuristic is to create a set of candidate pro-active caching points for each possible visited access router and then weighted by the probability of visiting each access router and explore node combinations for creating the service chain. 

\begin{enumerate}
\item For any request $r$, select the target node $d\in\mathbf{D}$ by highest $\rho_{d}$
and find the closest starting node $s\in \mathbf{S_{r}}$ by minimum shortest path routing cost $P_{sd}$;
\item On the shortest path from the selected $s$ and $d$, find all candidate nodes by $\mathbf{K}$;
\item Choose the closest $k$ from the selected $s$ on the path to host the NF$_{i}$ with the lowest visiting order sequence in request $r$ if there are enough resources to support the function, otherwise, host the sub-lowest function, until running out of resources;
\item Repeat step 2 and 3 until all NFs of request $r$ are hosted.
\end{enumerate}

\begin{algorithm}
\caption{PPCC}
 \SetKwInOut{Input}{Input}
 \SetKwInOut{Output}{Output}
\Input{$\mathbf{G}$; $\mathbf{D}$; $\mathbf{R}$; $\mathbf{K}$; $\mathbf{F}$; $\mathbf{H}$; attaching node $o$;}
\Output{VNF allocation: $x_{ri}^{k}$; PPCC cost: PPCC;}
PPCC$\leftarrow 0$\;
\For{$k \in \mathbf{K}$}{
Remaining utility of node $k$: $RU_{k} \leftarrow U_{k}$\;
}
Initialize all path bottlenecks: $R\Lambda \leftarrow \Lambda$\; 
\For{$i\in \mathbf{D}$ }{ 
	\If{$\rho_{i}==max(\rho_{i})$}{
		Destination node: $d\leftarrow i$\; 
    }
}
\For{$r\in \mathbf{R}$}{
     Starting node:$s \leftarrow$ find closest node s to d in $S_{r}$ with minimum $P_{sd}$\;
     candidate node priority list: $CPL \leftarrow \emptyset$\;
     $CPL \leftarrow $ sort $k\in$ \big\{ \{$n\vert n$ is on the shortest path from s to d\} $\cap$ $\mathbf{K}$ \big\} by the distance between $k$ and $s$ from low to high\;
     VNF priority list: $FPL \leftarrow \emptyset$\;
     $FPL \leftarrow $ sort $f_{i}$ by its visiting sequence $l$ of $r$\;
     former VNF location: $m \leftarrow s$ \;
     \For{$k\in CPL$}{
     	\For{$f_{i}$ $\in FPL$}{
        	\If{$u_{i} \leq RU_{k}$ \& \& $\lambda_{r} \leq R\Lambda_{km}$}{
            	host $f_{i}$ at $k$ \;
                $x_{ri}^{k} \leftarrow 1$ \;
                label $f_{i}$ as hosted at $k$ \;
                $RU_{k} \leftarrow RU_{k}-u_{i}$\;
                PPCC $\leftarrow$ PPCC$+C_{i}^{k}$\;
            }
        }
     }
     update $R\Lambda$;
}
\end{algorithm}

\section{Numerical Investigations}

\begin{table}[h]
\caption{Simulation Parameters}
\label{table1}
\begin{center}
\begin{tabular}{@{}lc@{}}
\toprule
\multicolumn{1}{c}{Parameter}                     & value      \\ \midrule
Number of candidate hosting nodes($K$)            & 20-50        \\
Degree per node                                   & 2-5        \\
Moving probability ($\rho_{d}$)                   & 0-1        \\
Number of starting points per request ($\mathbf{S_{r}}$)        & 1-5        \\
Number of destination points per user ($\mathbf{D}$)            & 1-5        \\
Number of requests per batch ($R$)                &50-200         \\
VNF number ($F$)                                  & 10         \\
Maximum number of VNF in a chain ($L$)            & 3-5         \\
Routing cost per link ($P_{km}$)                  & 1-100  \\
Cost per node to host VNF ($C_{i}^{k}$)            & 0      \\
Memory capacity per candidate node                & 8-16 GByte     \\
Number of virtual CPU cores per candidate node    & 32         \\
Memory requirement per VNF                        & 10-50 MByte \\
CPU core requirement per VNF                      & 0.125-0.25  \\
Flow rate requirement per chain ($\lambda_{r}$)                        &  0.064 - 10 Mbps \\
Capacity per link ($\Lambda_{km}$)                 & 2 Gbps
   \\ \bottomrule                                                 
                                                 
\end{tabular}
\end{center}
\end{table}

In this section, we provide a wide set of numerical investigations to evaluate the performance of proactive chaining-caching problem under various network scenarios.

We apply the proposed schemes on a wide set of random generated mobile edge networks. The applied networks are composed by a range of 20 to 50 candidate VNF hosting nodes and each candidate node has a degree ranged from 2 to 5. Besides, the number of starting points and destination points are set from 1 to 5. We assume that, the number of requests of a batch is from 50 to 200, and the number of different VNFs is 10.

As to the number of requests in the system, we assume that a total of 200 requests per second is generated and we take 0.25 to 1 seconds for each batch which converts to a total number of 50 to 200 requests per batch. We also assume that the flow rate requirement of a request varies from 64Kbps to 10Mbps. 

The moving probability to each destination node is randomly generated between 0 and 1, notice that, the summation of the moving probability to each destination of a mobile user does not exceed 1. With the aim to make VNF placement consider routing mostly, the VNF placement cost is set as 0. While, the shortest path routing cost can be any linear function of the traffic flows on the link, i.e., a delay, reliability, congestion, or energy metric. Without loss of generality, we choose a Open Shortest Path First (OSPF) or Enhanced Interior Gateway Routing Protocol (EIGRP) like routing metric in the simulation. To maintain the link diversity, we normalize the routing metric in the range from 1 to 100. In terms of physical resources of candidate VNF hosting node, we assume that each candidate node has 8 to 16 GByte memory capacity and 32 virtual CPU cores (e.g. a CPU with 8 cores 4 threads). While each VNF consumes memory in a range from 10 to 50 MBytes and uses 0.125 to 0.25 cores (i.e., each virtual CPU supports 4 to 8 VMs). As for the link capacity, we assume each link has a capacity of 2Gbps. The results are obtained by averaging 100 Monte Carlo simulations. To sum up, the parameters that have been used in the investigations are presented in Table \uppercase\expandafter{\romannumeral1}.

\begin{figure}
\centering
\includegraphics[width=0.76\columnwidth,trim=1cm 0cm 0cm 1cm]{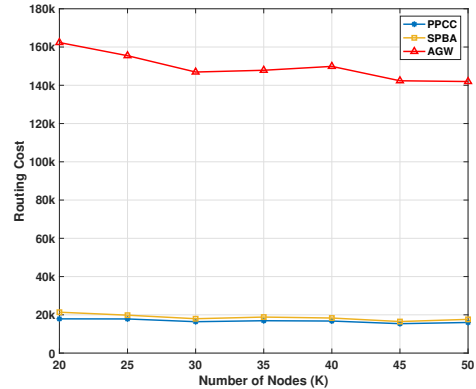}
\caption{Performance of the proposed scheme with different number of nodes in the network. (R=200)}
\label{fig:cost_no_nodes}
\end{figure}
\begin{figure}
\centering
\includegraphics[width=0.76\columnwidth,trim=1cm 0cm 0cm 2cm]{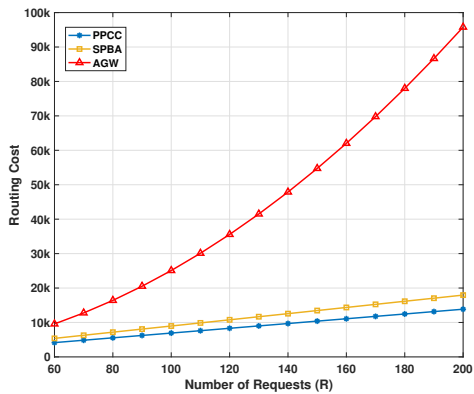}
\caption{Performance of the proposed scheme with increased number of service requests in the network. (K=20)}
\label{fig:cost_no_requests}
\end{figure}
\begin{figure}
\centering
\includegraphics[width=0.76\columnwidth,trim=1cm 0cm 0cm 1cm]{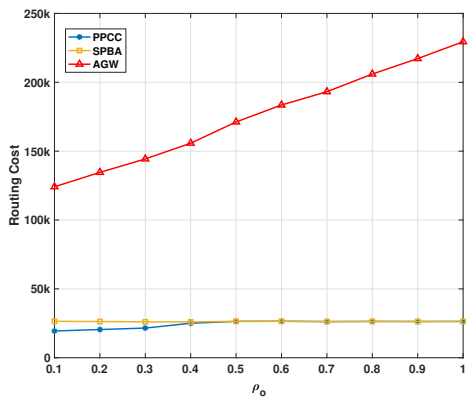}
\caption{Performance of the proposed scheme for different values of the parameter $\rho_0$. (K=20, R=200)}
\label{fig:cost_mobility}
\end{figure}

The proposed scheme is compared with two baseline schemes. In the first one, which provide a lower bound on the performance, content caching and VNFs are hosted at the network gateway, namely, AGW. The second scheme allocates caching as well as VNFs along the shortest path from the gateway node to the serving access router without considering mobility, and is called Shortest Path Based Allocation (SPBA). Figure \ref{fig:cost_no_nodes} shows the performance of the proposed scheme compared to the previous mentioned baseline techniques for different number of nodes in the network. As can be seen from the figure a performance gain of around $10\%$ for being mobility-aware (i.e., PPCC vs SPBA) can be achieved which is robust against different network sizes. This is justified by the fact that as the network resources are preserved for the most probable routing path, lower routing cost is obtained. A similar observation can be made from figure \ref{fig:cost_no_requests}, which shows the performance for different number of requests. With increased  number of requests, i.e., more constrained allocations, the performance gains increase from $22\%$ to $23\%$. It is noteworthy that, the performance metric shows a linear growth trend due to that each request are technically identical i.e., the number of VNFs in the chain is similar for each request. As a result, higher gain is expected in the case of dense mobile networks where the number of arriving requests in a certain amount of time is large. Finally, in figure \ref{fig:cost_mobility} we show the performance of the proposed scheme for different mobility use cases. The figure shows the performance gains as a factor of the parameter $\rho_{o}$. This parameter is defined as follows $\rho_{o} = 1 -  \sum_{d \in \mathbf{D}} \rho_d$, which means that as $\rho_{o}$ reaches close to 1 there is no mobility of the end-user, i.e., there is no change on the serving access router. As expected, there are no gains when there is no mobility, but as the mobility increase the gains reach more than $26\%$. The result suggests that, the proposed scheme is ideal for public transport mobile scenarios where mobile user ends will always leave for the next station.

%
%
\section{Conclusions}
In this paper, the rational of VNF location and chaining for proactive caching has been presented together with some key observations on this problem and the general principle of optimizing cache specific VNF service chains. Based on those preliminaries an optimization framework using integer linear mathematical programming has been detailed that integrates VNF chaining for proactive caching. In addition, since the problem resembles the UFL problem, which is $NP$-hard, a scale-free heuristic algorithm has been presented that can be applied in large network instances amenable for real time implementations. Finally, the attainable
performance of the proposed proactive caching service chains schemes was investigated.

\section*{Future Works}
We plan to carry out simulations on different network topology types (i.e., ring, star) along with detailed analyses on scalability and execution efficiency of the proposed schemes. We will provide comparisons between the optimal solution and heuristic solutions. End-user experience will be tested and proof of the NP-hardness of PCC problem will be given.


\end{document}